\documentclass[aps,preprintnumbers,prl,twocolumn,nofootinbib]{revtex4}

\usepackage{epsfig,latexsym,cancel,amssymb,amsmath,mathrsfs}
\usepackage{graphicx}
\usepackage{epstopdf}

\newcommand{\beq}{\begin{equation}}
\newcommand{\eeq}{\end{equation}}
\newcommand{\bea}{\begin{eqnarray}}
\newcommand{\eea}{\end{eqnarray}}

\begin{document}

\preprint{CALT-TH-2016-014}

\title{Strong CMB Constraint On $P$-Wave Annihilating Dark Matter}

 \author{Haipeng An}
\affiliation{Walter Burke Institute for Theoretical Physics,
California Institute of Technology, Pasadena, CA 91125}

\author{Mark B. Wise}
\affiliation{Walter Burke Institute for Theoretical Physics,
California Institute of Technology, Pasadena, CA 91125}

\author{Yue Zhang}
\affiliation{Walter Burke Institute for Theoretical Physics,
California Institute of Technology, Pasadena, CA 91125}


\begin{abstract}
We consider a dark sector consisting of dark matter that is a Dirac fermion and a scalar mediator. This model has been extensively studied in the past. If the scalar couples to the dark matter in a parity conserving manner then dark matter annihilation to two mediators is dominated by the $P$-wave channel and hence is suppressed at very low momentum. The indirect detection constraint from the anisotropy of the Cosmic Microwave Background  is usually thought to be absent in the model because of this suppression. In this letter we show that dark matter annihilation to bound states occurs through the $S$-wave and hence there is a constraint on the parameter space of the model from the Cosmic Microwave Background.
\end{abstract}

\maketitle

\noindent{\it Introduction.} The Standard Model (SM) has no acceptable dark matter (DM) candidate. As its name implies DM must be uncharged and various direct detection as well as astrophysical and cosmological constraints exist on its couplings to ordinary matter as well as its self interactions. These constraints motivate a class of very simple extensions of the SM that contain a dark sector with particles that carry no SM gauge quantum numbers. For thermal DM the minimal dark sector model consists of the DM and a mediator that the DM annihilates into in the early universe. There are various possibilities for the Lorentz quantum numbers of the DM and mediator. Two well studied examples are a Dirac fermion  with a mediator that is either a new massive $U(1)_D$ gauge boson (the dark photon) or a massive scalar. In the first case communication with the SM degrees of freedom occurs through the vector portal (via kinetic mixing between the $U(1)_D$ and $U(1)_Y$ field strength tensors) and in the latter case through the Higgs portal.

Constraints on the parameter space of these models occur from the so-called indirect detection signals. Annihilation of DM in the early universe at the time of recombination injects energy into the plasma of SM particles elongating the recombination process and changing expectations for the cosmic  microwave background (CMB) radiation anisotropy. Annihilation of DM today in our galaxy contributes to  electromagnetic and charged particle astrophysical spectra observed, for example, by the Fermi satellite. 

In a recent paper~\cite{An:2016gad}, we have highlighted the role that DM bound state formation can play on indirect detection signals from DM annihilation in our galaxy when the mediator is a dark photon (there bound state formation was not important for the CMB constraint). In this {\it letter}, we again consider the influence of DM bound state formation on indirect signals but focus on the case where the mediator is a real scalar and on the CMB constraint. 
We impose a parity symmetry on the dark sector with the real scalar mediator having even parity. Then, the Lagrange density for the DM sector is,
\beq\label{L}
{\cal L} =i\bar\chi\gamma^\mu\partial_\mu \chi - m_D \bar\chi \chi - g \bar\chi \chi \phi +\frac{1}{2}\partial_\mu\phi \partial^\mu\phi - \frac{1}{2} m_\phi^2 \phi^2 \ ,
\eeq
where $\chi$ and $\phi$ are the DM and the dark mediator and the Higgs portal couplings are omitted.
This model has been well studied for various reasons~\cite{Pospelov:2007mp, Kim:2008pp, Lin:2011gj, LopezHonorez:2012kv, Esch:2013rta, Kaplinghat:2013yxa, Franarin:2014yua, Wise:2014jva, Wise:2014ola, Kouvaris:2014uoa, Zhang:2015era, Shelton:2015aqa, Tsai:2015ugz, Krnjaic:2015mbs, Aravind:2015xst, Kim:2016csm, Bell:2016fqf}.
For DM heavier than $5-10$\,GeV, direct detection experiments~\cite{Akerib:2015rjg} and the requirement that $\phi$ decays before BBN set the lower bound, $m_\phi > 2 m_\mu \simeq 0.2\,$GeV. In our calculations below, we assume a thermal DM relic density, which fixes the value of $\alpha_D=g^2/(4 \pi)$ as a function of the DM mass, $m_D$.

%
%
%
%
%
%
%


\begin{figure}[t]
\centerline{\includegraphics[width=1\columnwidth]{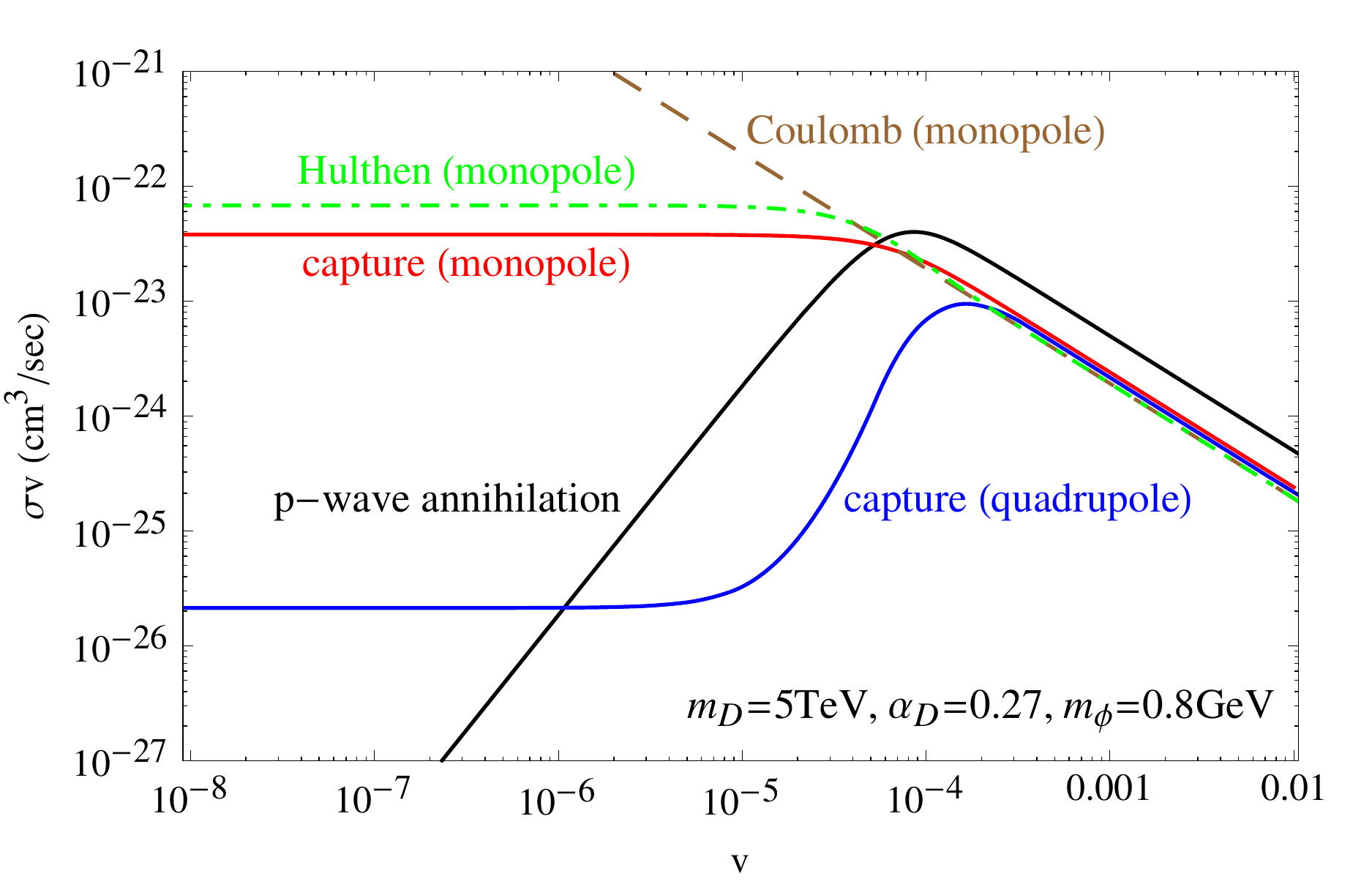}}
\vspace{-0.2cm}
\caption{DM relative velocity dependence in various cross sections. The black curve is the $p$-wave direct annihilation cross section for $\chi\bar\chi \to \phi\phi$. 
The red curve is the $(\chi\bar\chi)$ bound state formation cross section via monopole transition, evaluated numerically using Eqs.~(\ref{sigmavB}) and (\ref{monop}).
The blue curve stands for quadrupole transition counterpart.
The brown line is the monopole transition cross section in the Coulomb limit, while
the green curve is based on the Hulth\'en potential which gives a quite good approximation to the realistic Yukawa potential.\vspace{-0.4cm}}
\label{fig1}
\end{figure}

The most often considered DM annihilation process in this model is $\chi\bar\chi \to \phi\phi$. The parity of a $2\phi$ system must be even and so does the $\chi\bar\chi$ system because parity is conserved by the Lagrange density in Eq.~(\ref{L}). Therefore this annihilation is mostly $P$-wave for slow DM and anti-DM particles.\,\footnote{If parity was not conserved $S$-wave annihilation would be possible.}
With the $P$-wave Sommerfeld enhancement factor~\cite{Cassel:2009wt} included, the cross section times velocity can be written as
\beq
(\sigma v)_{\rm A}^{P\text{-wave}} = \frac{3\pi \alpha_D^2 v^2 }{8 m_D^2} \times \left|\sqrt{\frac{3}{4\pi p^2}}\frac{d R_{p1}}{ d r}(r=0)\right|^2 \ ,
\eeq
where $p = m_D v/2$, $v$ is the relative velocity. $R_{p\ell}$ is defined as the radial part of the initial scattering wave function (with the relative momentum aligned along the $z$-axis), $\Psi_{\vec p = p \hat z}(\vec r) = \sum_{\ell} R_{p\ell} (r) Y_{\ell 0}(\hat r)$, and $\Psi_{\vec p}(\vec r)$ is asymptotic to $\exp(i\vec p\cdot \vec r)$ at infinity. A typical curve of $(\sigma v)_{\rm A}^{P\text{-wave}} $ as a function of $v$ is the black curve in Fig.~\ref{fig1}. As $v$ gets smaller, $(\sigma v)_{\rm A}^{P\text{-wave}} $ first grows as $1/v$ due to the Sommerfeld enhancement, and then at around $v \sim m_\phi / m_D$, $(\sigma v)_{\rm A}^{P\text{-wave}} $ gets strongly suppressed. The drop-off is due to the {\it effective} potential barrier at $r \sim m_{\phi}^{-1}$ generated by the sum of the attractive Yukawa potential and the repulsive centrifugal potential. The transmission coefficient for tunneling through the barrier diminishes as $v^{2}$ in the small $v$ limit, as illustrated by Fig.~\ref{fig1}.

After thermal freeze out (chemical decoupling), DM can still maintain kinetic equilibrium with the $\phi$ particles in the universe. The DM velocity only red-shifts linearly with the expansion after the kinetic decoupling. For DM mass in the TeV range, their relative velocity $v$ during recombination is extremely small, $v \ll \sqrt{T_{\rm rec} /m_D} \sim 10^{-6}$, where $T_{\rm rec}$ is the temperature of the universe at the recombination era. Hence it has been thought that there will be no CMB constraint for the $P$-wave annihilating DM in this model. In this {\it letter}, we show that this is not the case. In some regions of parameter space, a pair of free DM particles can capture into a DM bound state via the emission of a $\phi$ particle, and then annihilate into $\phi$'s inside the bound state. The bound state formation process dominantly occurs in an $S$-wave and therefore is not suppressed at low velocity due to the absence of the centrifugal potential barrier. The mediator eventually decays to SM particles via the Higgs portal resulting in a CMB constraint on the region of the parameter space in the model where the kinematics allows for bound state formation.

\medskip
\noindent{\it Bound state formation cross section.}
The Hamiltonian for a non-relativistic DM-anti-DM system interacting with the mediator field is (in the center of mass frame) 
\begin{eqnarray}\label{hamilton}
H_{\rm int} &=& g \left[\phi\left({\vec r}/{2}\right)+ \phi\left(-{\vec r}/{2}\right) \right] \nonumber \\
&& - g \left[\phi\left({\vec r}/{2}\right)- \phi\left(-{\vec r}/{2}\right) \right] \frac{\nabla^2}{2m}  \ ,
\end{eqnarray}
where $g$ is the dark Yukawa coupling. $\vec r$ is the relative position of the DM-and-anti-DM particles, and $\phi$ is the Schr\"odinger picture mediator field.  In the bound state formation transition amplitude a mediator particle is created by the field $\phi$ . The mode expansion of the mediator field has exponential dependence on the wave-vector $\vec k$ that can be expanded, $ e^{\pm i {\vec k} \cdot {\vec r}} = 1 \pm  i \vec k \cdot \vec r - (\vec k \cdot \vec r)^2/2 + \cdots$. In the first line of Eq.~(\ref{hamilton}), due to the orthogonality between the initial and final states, the leading order contribution vanishes. The contributions at the $i\vec k \cdot \vec r$ order from DM and anti-DM cancel with each other. The contribution from the $(\vec k \cdot \vec r)^2$ order yields both monopole and quadrupole transitions. The second line of Eq.~(\ref{hamilton})  
represents the leading relativistic correction, which contributes to the monopole transition at the zeroth order in $\vec k\cdot \vec r$. 

The bound state formation cross section times the relative velocity can be written as
\beq\label{sigmavB}
\sigma v = \sum_f \sum_{\mathfrak{M}= M, Q} \int \frac{d^3\vec k}{(2\pi)^3 2k^0} (2\pi)\delta(E_f + k^0 - E_i) |V^\mathfrak{M}_{fi}|^2 \ ,
\eeq
where $E_i$ and $E_f$ are the energies of the initial and final states of the DM-anti-DM system. The sum over $f$ is over final bound state azimuthal, magnetic, and principal quantum numbers, but because we have aligned the dark matter relative momentum along the $z$-axis only the magnetic quantum number $m=0$ contributes.  Here we are neglecting the spin degrees of freedom for the dark matter. Including them would give a factor of $1/4$ from spin averaging and then for each $f=n,l,m$ there would be four final bound states; one with spin $0$ and three with spin $1$.

For the monopole (M) transition 
\beq
|V^M_{fi}|^2 \!=\! g^2 \left| \int dr r^2 \left[ \frac{1}{12} k^2 r^2 + \frac{\alpha_D e^{-m_\phi r}}{m_D r} \right] R_{n\ell}(r) R_{p\ell}(r)\right|^2 \ , \label{monop}
\eeq
where $k \equiv |\vec k|$, $R_{k\ell}$ and $R_{n\ell}$ are the initial and final radial wave functions. For quadrupole transition,
\begin{eqnarray}\label{quadrup}
|V^Q_{fi}|^2 \!\!&=&\!\!  \frac{g^2  k^4}{120}  \left[\frac{(\ell+1)(\ell+2)}{(2\ell+1)(2\ell+3)}\left|\int dr r^4 R_{n\ell}^*(r)R_{p\ell+2}(r)\right|^2\right. \nonumber\\
&&\!\!+\frac{2\ell(\ell+1)}{3(2\ell-1)(2\ell+3)} \left|\int dr r^4 R_{n\ell}^*(r)R_{p\ell}(r)\right|^2 \nonumber\\
&&\!\!\left.+ \frac{\ell(\ell-1)}{(2\ell-1)(2\ell+1)}  \left|\int dr r^4 R_{n\ell}^*(r)R_{p\ell-2}(r)\right|^2 \right]  .
\end{eqnarray}
During the time of recombination the DM and anti-DM particles have negligible kinetic energy, hence to emit an on-shell $\phi$, $m_\phi<\alpha_D^2m_D/(4n^2)$ is required in the Coulomb limit. This indicates $m_\phi \ll \alpha_D m_D/(2n)$. Therefore, the relevant bound state wave functions can be treated as Coulombic for the computation of the bound state formation cross section.
On the other hand, we solve for the scattering state wave functions numerically using the shooting method described in~\cite{An:2016gad}.

From numerical solutions, we find that after summing over the azimuthal quantum number $\ell$, for both the monopole and quadrupole transitions, $(\sigma v)\sim n^{-2}$ roughly. For $m_D = 5.0$ TeV, $\alpha_D = 0.27$, $m_\phi = 0.8$ GeV, the numerical solution of total cross sections times velocity for the monopole and quadrupole transitions are shown as the red and blue curves in Fig.~\ref{fig1} respectively. 

\begin{figure}[t]
\centerline{\includegraphics[width=1\columnwidth]{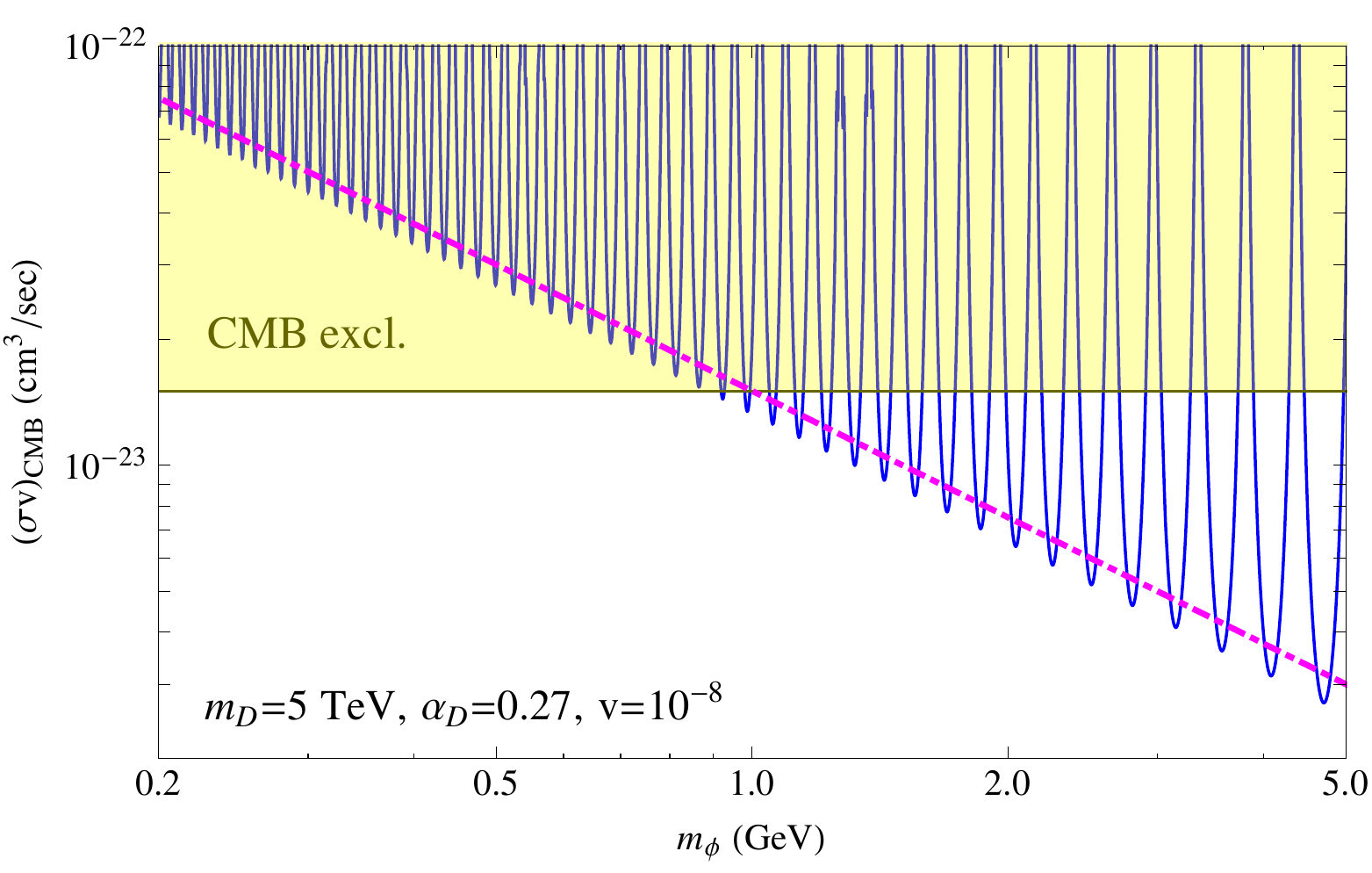}}
\caption{Scalar mediator mass dependence in the bound state formation cross section at very low DM velocity, $v\ll m_\phi/m_D$.
This is the cross section to be constrained by the CMB observation. In this plot, the yellow shaded region is excluded.
The magenta dot-dashed line is the appoximate envelop of the blue curves using Eq.~(\ref{11}).}
\label{fig2}
\end{figure}

For $v > m_\phi/m_D$, $\sigma v$ goes like $v^{-1}$  and  agrees with the result from the Coulomb potential scattering states which is shown by the brown line in Fig.~\ref{fig1}. 
For $v\ll \alpha_D$, the Coulomb scattering wavefunction takes the approximate form
\beq
R_{p\ell} (r) \simeq 4\pi \sqrt{\frac{2\ell+1}{4 p r}} J_{2\ell+1}\left( \sqrt{4 \alpha_D m_D r} \right) \ .
\eeq
In this limit, the monopole transition cross section times velocity can be written as
\beq
(\sigma v)^{M}_{n\ell} = \frac{2^{4\ell+7}(2\ell+1)n^{2\ell-2}\Gamma(n-\ell)\pi^2\alpha_D^5}{9  \Gamma(n+\ell +1) e^{4n} m_D^2 v}  \left(L_{n-\ell-1}^{2\ell+1}(4n)\right)^2 \ ,
\label{coulomb}
\eeq
where $L$ is the associated Laguerre polynomial, and here we have used Eq.~(7.421 (4)) of~\cite{book}. For the ground state $(n=1,\ \ell=0)$ formation, it can be simplified to $(\sigma v)^M_{10}=128\pi^2 \alpha_D^5/(9 e^4 m_D^2 v)$. The quadrupole piece is, $(\sigma v)^Q_{10}=512\pi^2 \alpha_D^5/(45 e^4 m_D^2 v)$. The $v^{-1}$ behavior originates from the Sommerfeld enhancement.

Due to the potential barrier, the contributions from incoming partial waves with $\ell > 0$ are suppressed when $v < m_\phi/m_D$. This causes the sharp drop-off in the smaller $v$ direction in the curve for quadrupole transitions in the region $v\lesssim m_\phi/m_D$, which is roughly $10^{-4}$ in Fig.~\ref{fig1}. The only transition  that does not get suppressed by the barrier is from $\ell = 0$ to $\ell=2$, which causes the blue curve to plateau at the small $v$ region. However, its value is suppressed by the phase space since the $\ell=2$ bound state starts from $n=3$.


In the case of finite $m_\phi$, the Huth\'en potential can be used as an approximation to the Yukawa potential. This is a useful approximation for $S$-wave scattering.
For  $m_\phi \ll \alpha_D m_D$, the incoming $S$-wave function can be approximated as ~\cite{Cassel:2009wt}
\beq
R_{p0} (r) = \sqrt \frac{4\pi}{\alpha_D m_D r} \left| \frac{\Gamma(a^-)\Gamma(a^+)}{\Gamma(1+2iw)} \right| J_{1}\left( \sqrt{4 \alpha_D m_D r} \right) \ ,
\eeq
where $w=m_D v/(2m_\phi)$, $a^{\pm} = 1+iw(1\pm\sqrt{1-x/w})$ and $x=2\alpha_D/v$.
One can get an analytic solution of the monopole transition into the $S$-wave bound state. In the limit $v\ll m_\phi/m_D$
\beq
(\sigma v)^M_{n0} = \frac{2^6\pi\alpha_D^4}{9n^3 m_D^2} \left| \frac{\Gamma(a^-)\Gamma(a^+)}{\Gamma(1+2iw)} \right|^2 e^{-4n}\left(L_{n-1}^{1}(4n)\right)^2  \ ,
\label{hulthen}
\eeq
This simplifies to,
\beq\label{11}
(\sigma v)^M_{n0} = \frac{2^6\pi^3\alpha_D^5 e^{-4n} \left(L_{n-1}^{1}(4n)\right)^2}{9n^3 m_D m_\phi \sin^2 \left( \pi\sqrt{\alpha_D m_D/m_\phi} \right)} \ ,
\eeq
unless the value of $\sqrt{\alpha_D m_D/m_\phi}$ is very close to an integer. The divergence one encounters in the cross section $(\sigma v)^M_{n0}$  using the expression above will be regularized by the small imaginary parts of $a^{\pm}$.
For values of $m_\phi$, where an $S$-wave state crosses threshold a peak appears in $(\sigma v)^M_{n0}$. This structure is depicted in Fig.~\ref{fig2}, where
the approximate lower envelop corresponding to Eq.~(\ref{11}) by sending the sine square factor in the dominator to unity is also shown as the magenta dot-dashed curve.

\medskip
\noindent{\it Annihilation decay.}
In the simple model of Eq.~(\ref{L}), there exist two ground states with quantum numbers $J^{PC}=1^{--}$ and $0^{-+}$. The $1^{--}$ state, once formed, is stable due to the $C$-parity symmetry. It is part of the dark matter. The $0^{-+}$ state, on the other hand, can decay. It is easy to verify that systems made of 2 and 3 real scalars are parity even. Because in this model, the Yukawa interaction also preserves parity, the leading decay channel of the $0^{-+}$ state is into $4\phi$'s, and the decay rate is,
\beq
\Gamma_{(0^{-+})\to4\phi} =\frac{F |\Psi_B(0)|^2 \alpha_D^4}{192 \pi^2 m_D^2}\ ,
\eeq
where $F\simeq 0.01$ has been determined numerically.

\medskip
\noindent{\it CMB constraint.}
The $0^{-+}$ bound state is spin-singlet and $1^{--}$, spin-triplet. Therefore, in this simple model (\ref{L}) only 1/4 of the dark bound states can decay. 
In the Coulomb limit, the ground state wave function at the origin is, $\Psi_B(0) = \sqrt{\alpha_D^3 m_D^3/(8\pi)}$. Thus, the lifetime of the $0^{-+}$ state is very short compared to the cosmological time scale during the recombination era. As a result, during recombination, the total formation rate of the $0^{-+}$ bound states is equal to their overall decay rate. 
The energy injection rate due to DM annihilation via the bound state channel is proportional to the $0^{-+}$ bound state formation cross section~\cite{An:2016gad}. (The  $1^{--}$ state is stable and these bound states are part of the DM today.)

\begin{figure}[t!]
\centerline{\includegraphics[width=1\columnwidth]{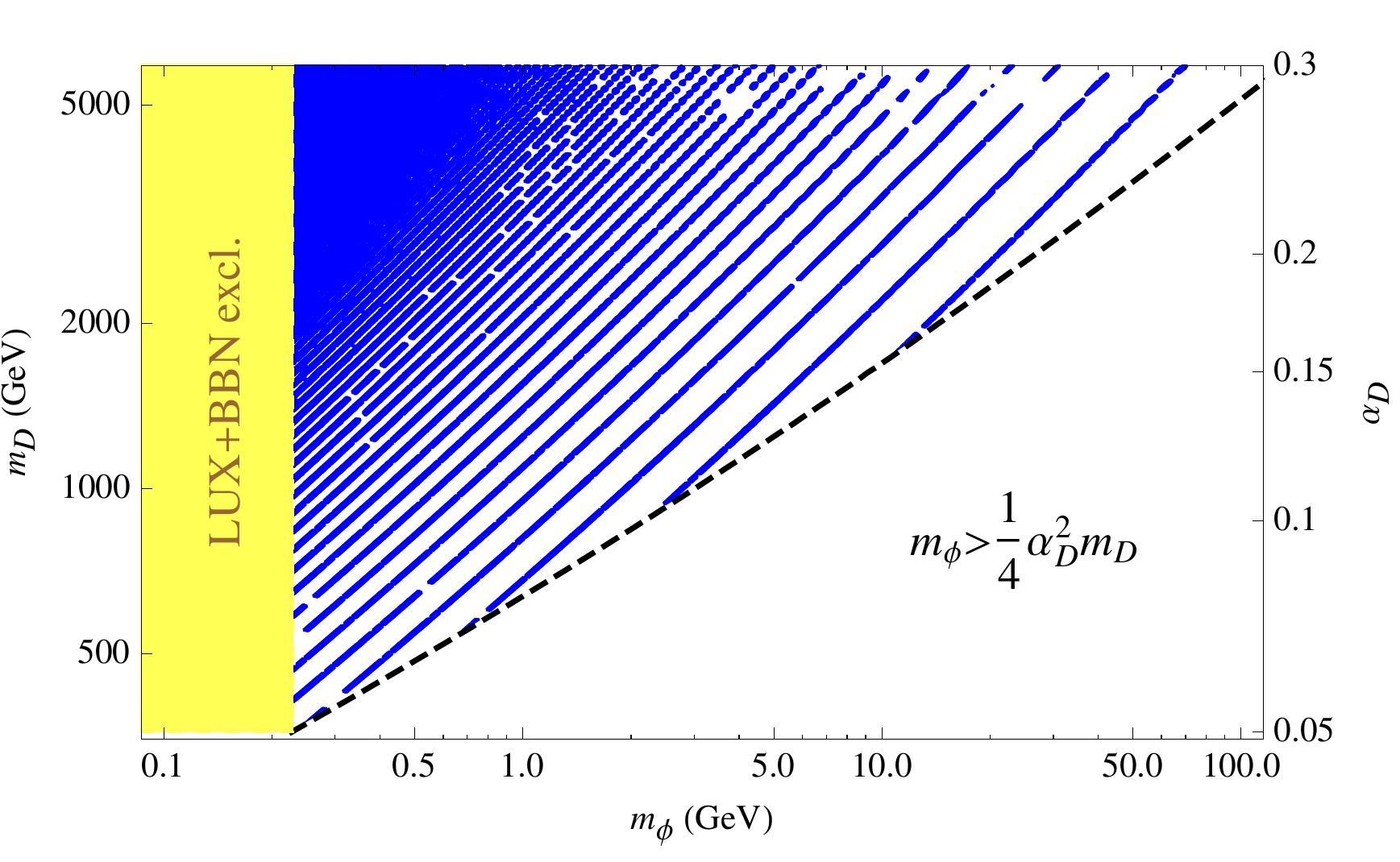}}
\caption{The blue region is the parameter space excluded by the CMB due to bound state formation, which is the main point of this work. The yellow region is known to be excluded jointly by BBN and direct detection experiments.}
\label{fig3}
\end{figure}

This cross section is bounded from above, in order not to distort the CMB spectrum, which is roughly~\cite{Slatyer:2015jla}, 
\beq
\lim_{v\to0} (\sigma v) < 3\times10^{-24} \,{\rm cm^3 sec^{-1}} \times \left( \frac{m_D}{\rm TeV} \right) \ .  
\eeq
Based on the bound state formation cross section we derived in Eq.~(\ref{sigmavB}), the CMB constraint is shown by Fig.~\ref{fig3}, where the blue region is excluded by current Planck data~\cite{Ade:2015xua}. The blue solid triangle region in the upper-left corner of Fig.~\ref{fig3} is fully excluded. The strips in the larger $m_\phi$ region are also ruled out due to the resonance effect.

\medskip
\noindent{\it Concluding Remark.} In this letter we have shown that for a scalar mediator dark matter annihilation into bound states  can give rise to qualitatively different physical effects. Without including bound state formation, at the time of recombination, the constraints in Fig.~\ref{fig3}  (which  are the main results of this letter) would be absent.

\medskip
\noindent{\it Additional Comments.}
The DM-anti-DM particles can also annihilate into $4\phi$'s when they are in an $S$-wave initial scattering state. With the Sommerfeld enhancement, the cross section times velocity for this channel is,
\beq
(\sigma v)_{\rm A}^{S\text{-wave}} = \frac{1}{4} \frac{|\Psi_S(0)|^2}{|\Psi_B(0)|^2} \Gamma_{(0^{-+})\to4\phi} \ ,
\eeq
where $|\Psi_S(0)|^2 = \left|{\Gamma(a^-)\Gamma(a^+)}/{\Gamma(1+2iw)}\right|^2$ is the scattering state wavefunction at the origin.
In practice, we find the ratio $(\sigma v)_{\rm A}^{S\text{-wave}}/(\sigma v)_{10}^M = 3e^4 F/(16384 \pi^3)\ll1$, so such a direct annihilation is numerically irrelevant throughout our analysis.



In Eq.~(\ref{L}) the DM is assumed to be a Dirac fermion. If it is a Majorana fermion, its direct annihilation is still dominated by the $P$-wave channel. In this case, due to its Majorana nature, the $S$-wave two-DM system, can only be in a spin singlet. As a result only 1/4 of DM annihilation can happen via the monopole transition into the ground state. One expects a similar constraint as in the case of Dirac DM. 

\medskip
\noindent{\it Acknowledgement.} 
We acknowledge an illuminating conversation with Haibo Yu. This work is supported by the DOE Grant DE-SC0011632, DE-SC0010255, and by the Gordon and Betty Moore Foundation through Grant No.~776 to the Caltech Moore Center for Theoretical Cosmology and Physics. We are also grateful for the support provided by the Walter Burke Institute for Theoretical Physics.

\end{document}